\documentclass[3p,times,twocolumn]{elsarticle}

\usepackage{ecrc}

\volume{00}

\firstpage{1}

\journalname{Nuclear Physics B Proceedings Supplement}

\runauth{R. D. Blandford et al.}

\jid{nuphbp}

\jnltitlelogo{Nuclear Physics B Proceedings Supplement}

\usepackage{amssymb}

\usepackage[figuresright]{rotating}

\usepackage{amsmath}
\usepackage{xspace}

\begin{document}

\begin{frontmatter}



\dochead{}

\title{Cosmic Ray Origins: An Introduction}


\author[label1]{Roger Blandford}
\author[label1]{Paul Simeon}
\author[label1]{Yajie Yuan}

\address[label1]{KIPAC, Stanford University}

\begin{abstract}
Physicists have pondered the origin of cosmic rays for over a hundred years. However the last few years have seen an upsurge in the observation, progress in the theory and a genuine increase in the importance attached to the topic due to its intimate connection to the indirect detection of evidence for dark matter. The intent of this talk is to set the stage for the meeting by reviewing some of the basic features of the entire cosmic ray spectrum from GeV to ZeV energy and some of the models that have been developed. The connection will also be made to recent developments in understanding general astrophysical particle acceleration in pulsar wind nebulae, relativistic jets and gamma ray bursts. The prospects for future discoveries, which may elucidate the origin of cosmic rays, are bright.
\end{abstract}


\end{frontmatter}


\section{The Accelerating Universe}
\label{sec:observations}
\subsection{Direct Measurements}
\label{subsec:direct}
\begin{figure}
  \includegraphics[width=0.45\textwidth]{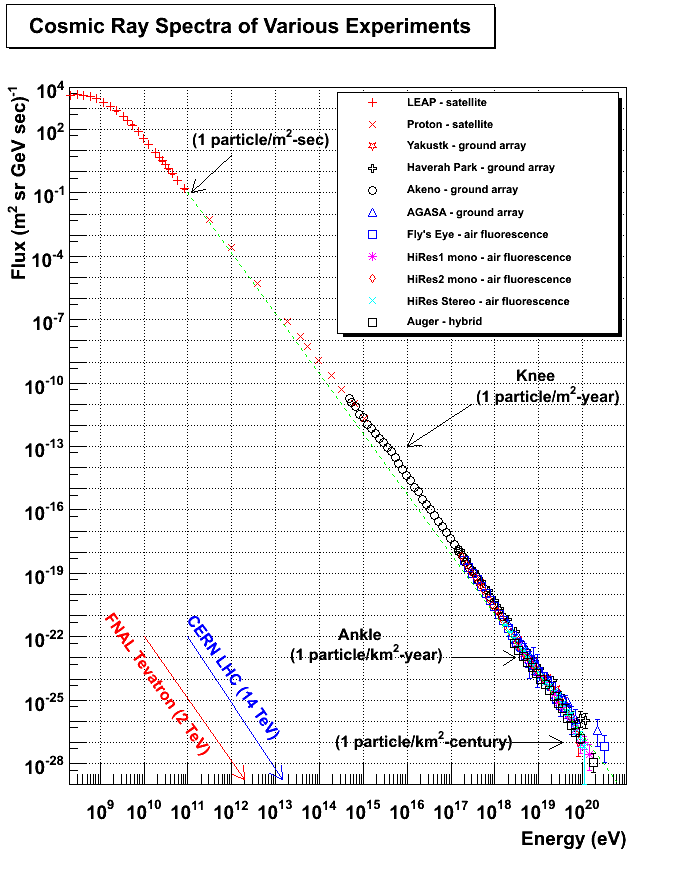}\\
  \caption{Overall energy spectra of cosmic rays from various experiments. (Courtesy of Dr William Hanlon.)}\label{fig:CRspectrum}
\end{figure}
\begin{figure}[htb]
  \includegraphics[width=0.4\textwidth]{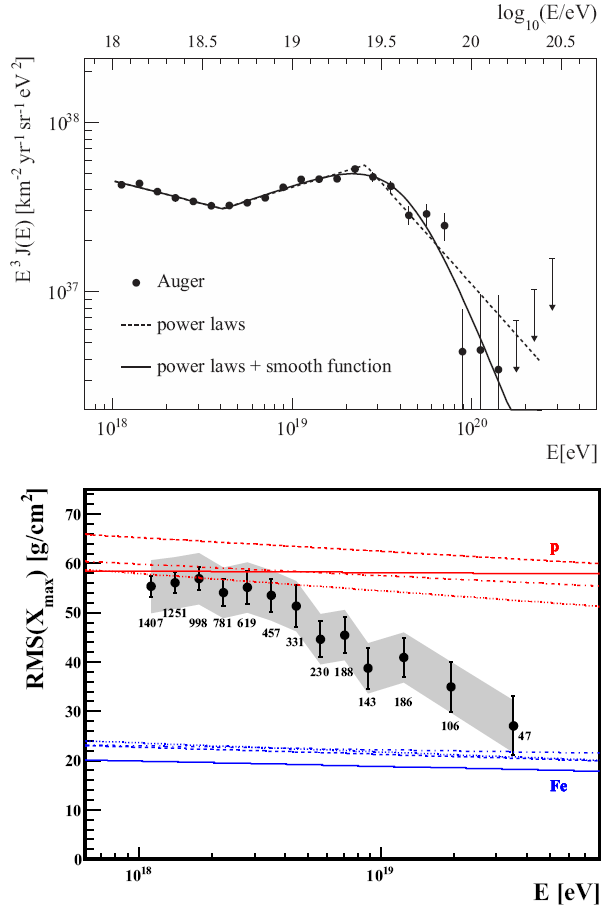}\\
  \caption{Spectrum of UHECR (top panel) and fluctuation of the atmospheric depth $X_{max}$ around its mean value $RMS(X_{max})$ as a function of energy (bottom panel) measured by Pierre Auger Observatory (from \cite{deMelloNeto2014AdSpR..53.1476D}).}\label{fig:PAO_spectrum_composition}
\end{figure}

The observed cosmic ray spectrum extends from $\sim$~MeV to $\sim$~ZeV\footnote{Almost the energy of a Marchisio goal, but the momentum of a snail!} energy --- fifteen decades in total (Figure \ref{fig:CRspectrum}). The lowest energy particles derive from the sun, the solar wind and the planets and provide (thankfully) small scale illustrations of general mechanisms that operate in the interstellar and intergalactic media. In particular, the two faithful Voyager spacecraft have traversed the solar wind termination shocks, and one of them appears to have crossed the heliopause and is now sampling interstellar cosmic rays directly \cite{Stone2013Sci...341..150S,Krimigis2013Sci...341..144K}. The $\sim$~GeV -- $\sim100$~TeV particles, mostly come form Galactic supernova remnants (e.g., \cite{Drury2012APh....39...52D, Zirakashvili2013JPhCS.409a2012Z, Blasi2013A&ARv..21...70B}). Recent advances include the consistent measurement of their age at low energy and the measurement of subtle features in their energy spectra by experiments like CREAM \cite{Seo2014AdSpR..53.1451S}. The particles between the ``knee(s)'' and the ``ankle'' have a changing composition and may be the heavy counterparts of the highest rigidity protons (e.g., \cite{KASCADE-Grande2013PhRvD..87h1101A}; and \cite{Blasi2014arXiv1403.2967B} for a review). Larger shocks than those associated with individual supernova remnants, perhaps supershells or a Galactic wind termination shock are suggested. Alternatively, neutron stars in the form of pulsars (and their surrounding wind nebulae) are credible sources. The highest energy particles  --- Ultra High Energy Cosmic Rays (UHECR) --- appear to manifest the ``GZK'' cutoff due to photo-pion production on the microwave background and seem to have a composition that is changing from hydrogen to iron as shown in Figure \ref{fig:PAO_spectrum_composition} \cite{PAO2010PhRvL.104i1101A, deMelloNeto2014AdSpR..53.1476D, Sokolsky2013EPJWC..5206002S}, perhaps for the same reason as the $\sim$~PeV particles. Proposed sources include Active Galactic Nuclei (AGN, e.g., \cite{Berezhko2008ApJ...684L..69B}), Gamma Ray Bursts (GRB, e.g., \cite{Meszaro2007auhe.conf...97M}), spinning magnetars (e.g., \cite{Fang2012ApJ...750..118F}), and intergalactic shock fronts (e.g. \cite{Kang1996ApJ...456..422K}).

\subsection{Indirect Observations}
\label{subsec:indirect}
\begin{figure*}[htb]
  \centering
  \includegraphics[width=0.8\textwidth]{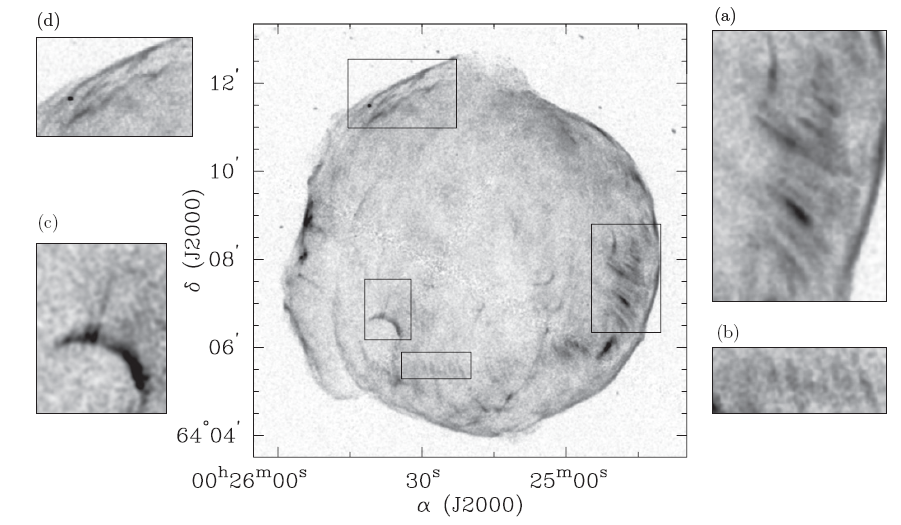}\\
  \caption{\emph{Chandra} X-ray $4.0-6.0$~keV image of the \emph{Tycho} supernova remnant, showing detailed non-thermal emitting features manifesting the shock fronts (from \cite{Eriksen2011ApJ...728L..28E}).}\label{fig:Tycho}
\end{figure*}

\begin{figure}[htb]
  \includegraphics[width=0.45\textwidth]{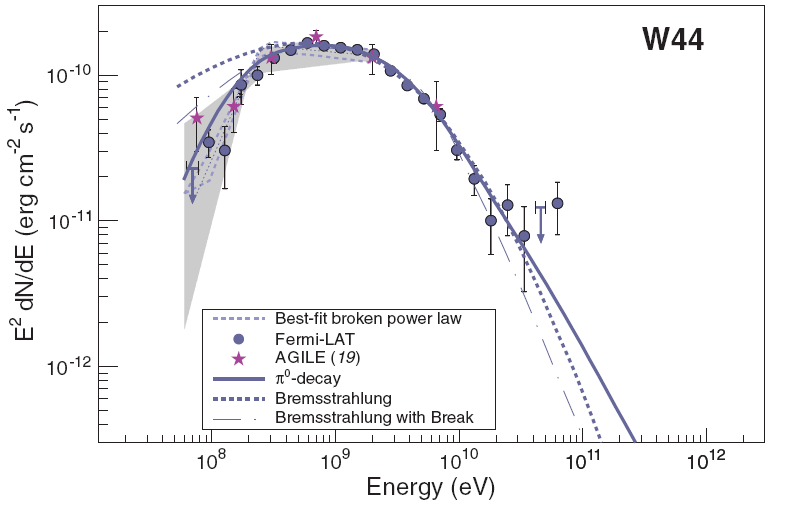}\\
  \caption{Gamma-ray spectrum of W44 as measured with the Fermi LAT, which shows good agreement with pion-decay gamma-ray production model (from \cite{Pion2013Sci...339..807A}).}\label{fig:W44}
\end{figure}

Of course most of what we know about putative cosmic ray sources derives from electromagnetic observations throughout the $\sim70$ octaves open to astronomical observations. The Chandra X-ray Observatory (CXO) has nearly matched the resolution of radio maps of SuperNova Remnants (SNR) and exhibited electron synchrotron emission from the bounding shock fronts (Figure \ref{fig:Tycho}). They have also produced compelling evidence that high Mach number shocks stretch and amplify magnetic field as well as accelerate cosmic ray electrons with energy up to $\sim100$~TeV \cite{Eriksen2011ApJ...728L..28E}. While, it was hard to doubt that this was accompanied by proton acceleration, direct evidence has only been presented recently by \emph{Fermi} and AGILE observations which exhibit the predicted ``pion bump'' in the $\gamma$-ray observations in middle-aged supernova remnants expanding into dense molecular gas \cite{Pion2013Sci...339..807A, AGILE_W44_2011ApJ...742L..30G} (Figure \ref{fig:W44}). No less dramatic have been the observations by Atmospheric Cerenkov Telescopes (ACTs) of TeV $\gamma$-rays which show evidence for efficient acceleration beyond $\sim100$~TeV (e.g., \cite{Aharonian2007A&A...464..235A, Aharonian2007ApJ...661..236A}).

CXO and XMM-Newton observations of clusters of galaxies have demonstrated that the entropy of gas in the outskirts of rich clusters of galaxies is extremely high \cite{Urban2014MNRAS.437.3939U} and is strongly suggestive of the presence of high Mach number accretion shocks\footnote{Radio features that have been identified with giant shocks may not have high enough Mach numbers to be efficient particle accelerators} as had been predicted on the basis of cosmological simulations \cite{Skillman2008ApJ...689.1063S}. A second development is the measurement of the iron abundance which can be as high as one third solar \cite{Werner2013Natur.502..656W}. These two observations support the intergalactic shock explanation of UHECR.

Part of the reason for a resurgence of interest in cosmic ray astrophysics is the race to identify dark matter. As is well known, the most compelling candidate is one or more new ``Weakly Interacting Massive Particles'' (WIMPs) that have been postulated to be the supersymmetric partners to the particles of the standard model. Exquisite experiments below ground (in mines, e.g. \cite{CDMS2013PhRvL.111y1301A}), on ground (at the Large Hadron Collider \cite{ATLAS2013JHEP...04..075A, CMS2012JHEP...09..094C}) and above ground (using \emph{Fermi} and other $\gamma$-ray telescopes \cite{Funk2013arXiv1310.2695F}) have each furnished impressive constraints on the properties of the presumptive particles. Cosmic ray physics is the mother of particle physics and having produced the first detections of positrons, pions, muons and kaons could be the first to detect another radically new class. However, no such claim should ever be accepted ahead of a confident understanding of the more prosaic and conventional cosmic ray foregrounds and backgrounds.
\subsection{Crab Nebula}
\label{subsec:crab}
\begin{figure}[htb]
  \includegraphics[width=0.45\textwidth]{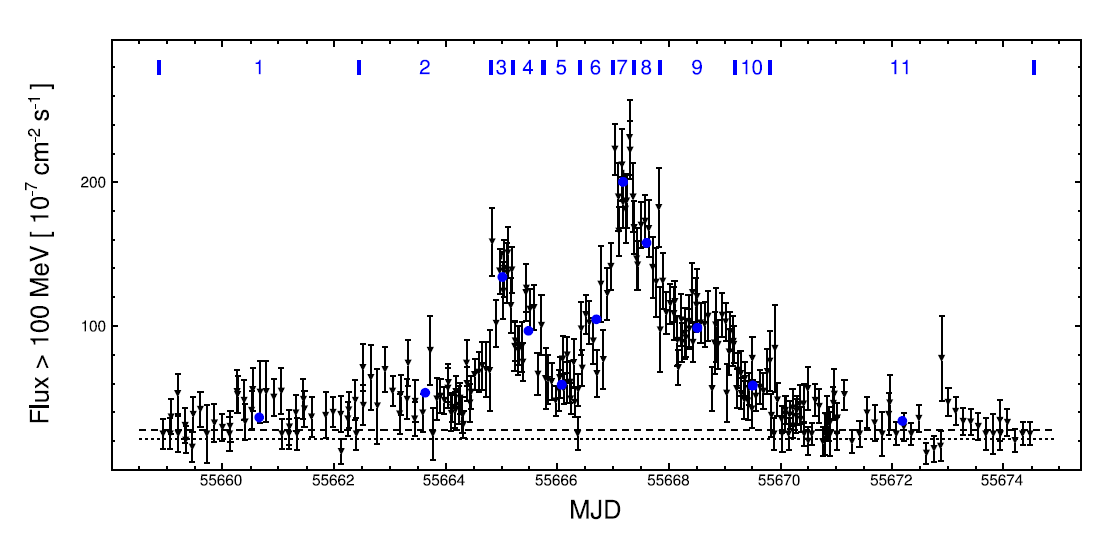}\\
  \caption{Integrated flux above 100 MeV from the Crab Nebula and pulsar as a function of time during the 2011 April flare (from \cite{Buhler2014RPPh...77f6901B}).}\label{fig:Crab_lightcurve}
\end{figure}

\begin{figure}[htb]
  \includegraphics[width=0.45\textwidth]{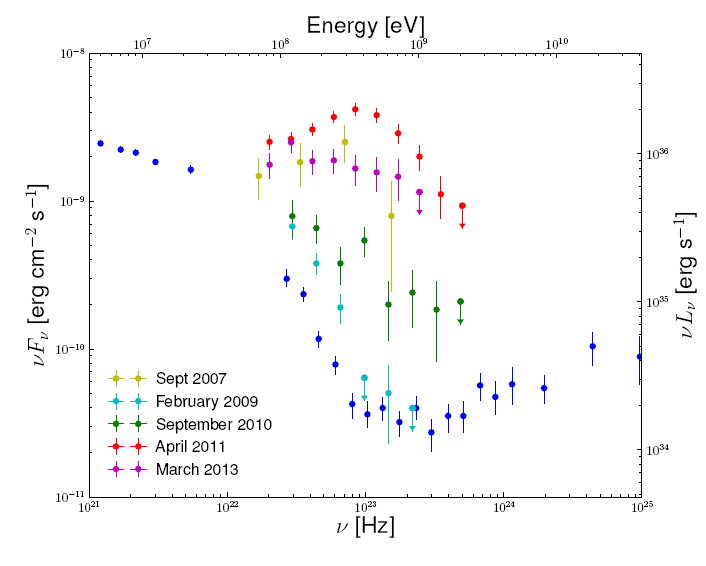}\\
  \caption{Spectral energy distribution at the maximum flux level for five of the Crab Nebula flares. Average nebula flux is shown as the blue points. (From \cite{Buhler2014RPPh...77f6901B}).}\label{fig:Crab_flare_spectra}
\end{figure}

The Crab Nebula has long been our best high energy astrophysics laboratory and many common effects have first been identified there. It is not disappointing us. Recent discoveries include $\sim400$~GeV pulsation \cite{VERITAS_Crab_pulsar2011Sci...334...69V,MAGIC_Crab_pulsar2012A&A...540A..69A}, rapid secular variation in the total nebular flux \cite{Wilson-Hodge2011ApJ...727L..40W} and rapid variation of the ``inner knot'' which may mark the termination of the wind in the inner nebula \cite{Moran2013MNRAS.433.2564M}. However the most striking discovery, which may presage a new kind of particle acceleration with clear relevance to cosmic ray origin, is the discovery of dramatic, $\sim10$~hr $\gamma$-ray flares which are localized around 400~MeV and are not obviously seen in any other spectral band (eg. \cite{AGILECrab2011Sci...331..736T,FermiCrab2011Sci...331..739A} and \cite{Buhler2014RPPh...77f6901B} for a review). Figure \ref{fig:Crab_lightcurve} shows the light curve of the biggest flare in April 2011. The flux doubled within $t_d\lesssim8$ hours at the rising edges. The spectral energy distribution for a few flares taken near the peak flux level is shown in Figure \ref{fig:Crab_flare_spectra}. The emission process is most likely to be synchrotron, requiring electrons/positrons of energy $\sim3$~PeV in a $\sim1$~mG magnetic field. A peak of $\sim400$~MeV indicates very efficient acceleration that goes beyond the classical radiation reaction limit in an MHD setting. These flares are not accompanied by changes in the pulsar timing and so presumably originate in the nebula which is many light years in size. The isotropic energy radiated in the strongest flare was $\sim10^{34}$~J which is equivalent to the energy contained within a region about a hundred, not ten, light hours across. Either there is strong relativistic beaming or some way must be found to concentrate energy within a small volume (or both).

\subsection{Relativistic Jets}
\label{subsec:reljets}
\begin{figure}[htb]
  \includegraphics[width=0.45\textwidth]{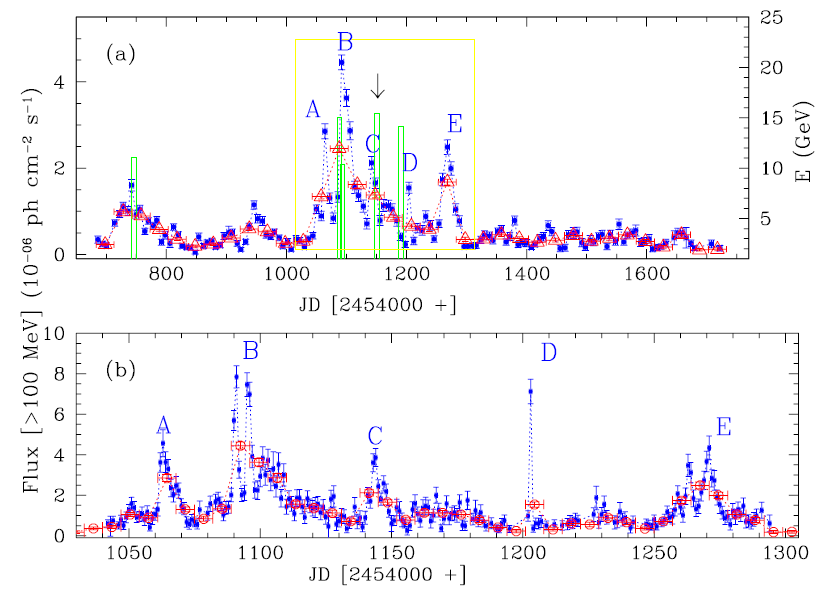}\\
  \caption{(a) Weekly (blue) and monthly (red) averaged flux ($E>100$~MeV) light curve of 3C 273 measured by \emph{Fermi}-LAT. Green histogram represents the arrival time distribution of $E>10$~GeV photons. (b) One-day averaged light curve during the active period within the yellow box of (a). From \cite{Rani2013A&A...557A..71R}.}\label{fig:3C273}
\end{figure}

\emph{Fermi} and the ACTs have also made dramatic observations of relativistic jets from AGN, GRB and binary sources. Blazars (AGN directed towards us) exhibit variability on timescales that can be as short as few minutes (e.g., \cite{Aleksic2011ApJ...730L...8A,Fortson2012AIPC.1505..514F}). However, we can place a lower bound on the radius of emission because the gamma rays have to avoid pair production as they escape the near infrared photons. The far more luminous quasars, for example 3C273\footnote{$\sim15$~GeV photons have been observed varying in half a day from a radius that must be at least 30 light years}, also exhibit this phenomenon on a larger and slower scale \cite{Rani2013A&A...557A..71R}, see Figure \ref{fig:3C273}. Similar constraints are imposed by \emph{Fermi} observations of high energy photons from GRBs. Again, the inference from many jet sources is that particles can be promptly accelerated with impressive efficiency in tiny magnetized regions within extended sources.

\subsection{VHE Neutrinos}
\label{subsec:vhenus}
\begin{figure}[htb]
  \includegraphics[width=0.45\textwidth]{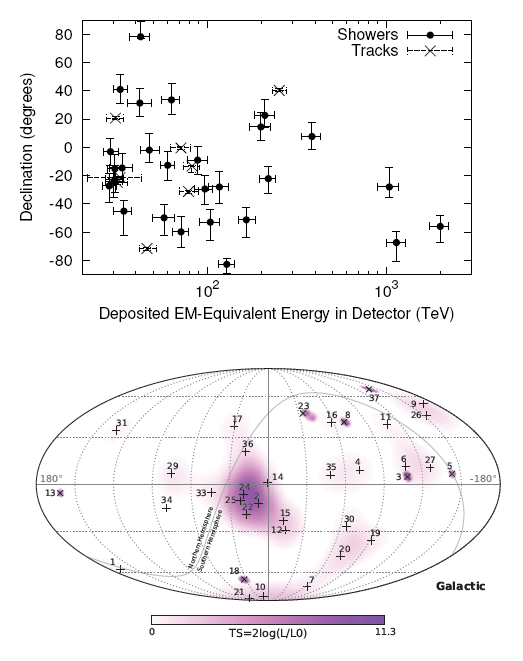}\\
  \caption{Deposited energies of the VHE neutrinos and their arrival directions (from \cite{IceCube2014arXiv1405.5303A}).}\label{fig:neutrino}
\end{figure}

Another recent development has been the beginning of VHE neutrino astronomy. As of this writing, 37 neutrinos have been reported by IceCube with energy that can exceed a few PeV \cite{IceCube2014arXiv1405.5303A} (Figure \ref{fig:neutrino}). They are argued to have a cosmic origin but no identification --- spatial or temporal --- have been found and they are not significantly clustered.

\subsection{Fast Radio Bursts}
\label{subsec:frbs}
The most recent phenomenon which has an even more tantalizing observational status is the strong evidence that there is a class of cosmologically distant source that produce highly coherent, millisecond radio bursts with an isotopic energy $\sim10^{32}$~J every five minutes \cite{Thornton2013Sci...341...53T}.

What this brief overview demonstrates is that we are still in the discovery epoch of high energy astrophysics and that new sources with new mechanisms may be accelerating familiar and unfamiliar populations of high energy particles.
\section{General Inferences and Principles}
\label{sec:general}
\subsection{Efficiency}
\label{subsec:efficiency}
When trying to explain these observations, it is helpful to keep in mind some general inferences and principles. The first is that the most dramatic feats of cosmic particle acceleration are impressively efficient. SNR, AGN, GRB and so on seem to be diverting at least ten percent of the available free energy into suprathermal particles. It is often hard to be quantitative because we are often hostage to poorly constrained astrophysical measurements and features like relativistic beaming may cause us to exaggerate the nonthermal power, but there are several cases where this statement is secure. The concentration of power into extremely suprathermal particles is typical. For example, some SNR channel one percent of the available energy into particles with individual energy $\sim10^{10}$ the mean energy per particle at the shock front. (If UHECR originate in intergalactic shocks a comparable fraction of the power goes into particles that are $\sim10^{18}$ times thermal energies. Greed is very good to UHECRs!)
\subsection{Current}
\label{subset:current}
Now this astonishing efficiency has a seemingly inescapable physics consequence. Only electric field can do work on a charged particle and change its energy; magnetic field is ``lazy''. Direct electrostatic acceleration is observed in terrestrial aurorae and is surely occurring in pulsar magnetospheres. Under electrostatic conditions a potential difference of 1~ZV is needed to accelerate a proton to 1~ZeV even if the particle follows a wildly gyrating path to cross it. Of course, if the electromagnetic field is time-dependent a small minority of particles can follow improbable paths to keep gaining energy, but large potential differences are still necessary. Now, the power that goes into the high energy particle is the product of the electrical current that they carry and the potential difference. Put another way, efficiently-accelerated particles must be a significant source for the electromagnetic field and an agency for its rapid change. The Fermi-inspired picture of particles rattling around a cosmic pinball machine is incomplete.

These considerations force one to think seriously about the current flow as a complement to a description of the magnetic field in the source. Now electrical current can act in two distinct ways on plasma. It can dissipate -- heating up the particles at the expense of the electromagnetic energy. We might expect that the energy would be shared between the majority of the particle, but the evidence points to just the opposite behavior being common in highly collisionless plasma\footnote{Collisional plasmas must quickly share the energy as the Coulomb cross section increases with decreasing frequency. The ``collisions'' responsible for the dissipation in a ``collisionless'' plasma are with large scale electromagnetic field as happens with synchrotron radiation, Compton scattering or wave-particle interactions.}. We seek acceleration mechanisms that allow a few particles to run away with most of the energy. This presents a challenge and a clue.

The second type of current-plasma interaction is through the Lorentz force which can do work without creating entropy\footnote{We might call this action ``industrious'', in contrast to the ``prodigal'' action when dissipation is present.}. The result will be the transformation of electromagnetic power into kinetic energy. Many particle acceleration schemes do just this and then accelerate particles and re-create magnetic field downstream at a high Mach number shock.
\subsection{Escape and Propagation}
\label{subsec:propagation}
Another important consideration, to which we have already alluded, is that the particles or the photons they emit must be able to escape the acceleration site. The highest energy cosmic rays can only propagate $\sim30$~Mpc in the darkest and lowest density region in the universe. Conditions are much more restrictive around AGN and GRB! Electrons and positrons are especially susceptible to radiative loss and they cannot be carrying the momentum in a relativistic jet all the way from a spinning black hole to radii where the jet is actually observed. Electromagnetic field, or possibly ions must do the job. (Neutron-based schemes require that the neutron, which has a half life of $\sim10$~ min in its rest frame, is created with sufficient energy that it can escape through time dilation. This is a challenging constraint.)

The longest-studied propagation problem is that of Galactic cosmic rays. Here, the simple theory is that relativistic protons (plus heavier nuclei and electrons) are created (by supernova remnants) with a source spectrum $S(E)\propto E^{-2.2}$ and that their residence time in the Galaxy, inferred from the observed energy-dependence of the fluxes of secondary nuclei,  like Li, Be, B, to those of their primary cosmic ray parents, such as C, N, O, varies $\propto E^{-0.5}$. Combining this with the source spectrum leads to a mean primary cosmic ray flux $\propto E^{-2.7}$. The cosmic ray energy density is dominated by mildly-relativistic, $\sim1$~GeV protons and is comparable with the interstellar energy densities associated with thermal gas, magnetic field, stellar radiation and the cosmic microwave background. Combining the secondary to primary ratio of these particles with the known spallation cross sections in collisions with stationary protons gives an estimate of the mean grammage traversed by these particles before they escape the Galaxy $\sim6$~g cm$^{-2}$. Now, we know that the mean grammage of interstellar gas through the disk of the Galaxy at the solar neighborhood  is $\sim2$~mg cm$^{-2}$ and so a typical GeV cosmic ray traverses the galactic disk roughly 3000 times allowing us to estimate the Galactic cosmic ray power as $\sim3\times10^{33}$~W or $\sim0.03$ of the supernova power assuming a Galactic supernova rate of one per century. Measuring the fraction of the radioactive isotope $^{10}$Be allows one to estimate the mean residence time of these cosmic rays in the Galaxy of $\sim20$ million years, implying that they occupy a disk with scale height about five times that of the gas. Now none of these quantities is well-defined, let alone well-measured, but the arithmetic is broadly consistent with more modern simulations using codes like GALPROP \cite{Moskalenko2012cosp...39.1281M} which point to a somewhat higher efficiency as mentioned above.

One effect that might be important for the interpretation of the excess positron cosmic ray fraction reported by PAMELA \cite{PAMELA2009Natur.458..607A}, \emph{Fermi} \cite{positronFermi2012PhRvL.108a1103A} and AMS \cite{AMS2013PhRvL.110n1102A} is that the positrons are scattered by the same Alfv\'{e}n waves as protons of the same rigidity. Electrons are scattered by different waves. As these waves are created by the protons and electrons and there are more of the former than the latter, the positrons might be retained by the Galaxy for longer and consequently be over-represented in the measured spectrum without actually needing to invoke a pulsar or dark matter source.

\section{Acceleration Mechanisms}
\label{sec:acceleration_mechanisms}
\subsection{Unipolar Inductors}
\label{subsec:unipolar}

One type of accelerator is the unipolar inductor, more or less as first envisaged by Faraday. A spinning conductor with angular frequency $\Omega$ and threaded by magnetic flux $\Phi$ is capable of generating an EMF $V\sim\Omega\Phi\sim E_{max}/e$ \cite{Goldreich1969ApJ...157..869G}. Under electromagnetic conditions which are optimal, the effective impedance of the source region is roughly that of free space $Z\sim100\Omega$ allowing us to estimate the minimum current and power. UHECR accelerated in this way require a power of at least $\sim10^{39}$~W $\propto E_{max}^2$ which cuts down the options. The two candidate sources are neutron stars and black holes. It is impressive that leading models of GRBs are in principle capable of producing $\sim100$~ZV of EMF and $\sim1$~ZA of current and this makes them appealing sources. The challenge is to explain how UHECR can escape the ``brightest objects in the universe'' and survive the long journey to Earth.

There has been much impressive theoretical progress in understanding neutron stars and black holes. The electromagnetic structure of a force-free pulsar magnetosphere, which was once seen as a straightforward exercise in electrical engineering, has only recently yielded to numerical assault and the solution is still sensitive to the detailed description of the current sheet boundaries (e.g., \cite{Spitkovsky2006ApJ...648L..51S}). This research combined, with $\gamma$-ray observations, has suggested new answers to the question of ``How do pulsars shine?'' but they have also motivated a new description of pulsar winds (e.g., \cite{Arons2012SSRv..173..341A}). The old viewpoint was that the power left the pulsar surface almost totally in electromagnetic (Poynting flux) form and that somewhere, presumably beyond the light cylinder at least 99.9 percent of this outflowing electromagnetic power is invisibly (industriously not prodigally) \footnote{The big surprise about the Crab inner knot mentioned above  is that so \emph{little} wind power is radiated.} converted into relativistically outflowing plasma which then accelerates particles at a termination shock analogous to the solar wind termination shock \cite{ReesGunn1974MNRAS.167....1R, KC1984ApJ...283..694K, KC1984ApJ...283..710K}. The magnetic structure of the wind is mostly toroidal and comprises an AC (striped) part at low latitude and a DC part, roughly $\propto\sin\theta/r$ that changes sign between the two hemispheres. The modern view buttressed by simulation is that the outflow is closer to equipartition and that large loops of toroidal field are supplied to the nebula which must be converted to relativistic particles within the nebula and that somehow this powers the particle acceleration and synchrotron radiation from the nebula (e.g., \cite{Porth2014MNRAS.438..278P}). Using the simple arithmetic from the last paragraph, the Crab Pulsar is thought to produce an EMF $\sim50$~PV and a power $\sim3\times10^{31}$~W, comparable with that radiated.

Astrophysical black holes can also behave in this fashion. They are described by the Kerr\cite{Kerr1963PhysRevLett.11.237} metric and measured by their mass, $m$, which also provides a scale for the energy, length, time , power \dots, and the spin $\Omega$ which parametrizes the curvature off the surrounding spacetime and is also a measure of how much extractable, rotational energy is associated with the black hole. A maximally spinning hole has $\sim0.3$ of its mass available in principle. The best way to extract this mass by threading the event horizon with magnetic field supported by external current and anchored by the inertia of a surrounding (thick) accretion disk \cite{BZ1977MNRAS.179..433B}. The black hole is acting like a unipolar inductor with the important difference from a neutron star that it possesses a significant internal resistance $\sim100\Omega$. So a large scale electrical circuit with the spinning black hole behaving as a battery will only dissipate roughly half of the available energy beyond the event horizon. Typically $\sim0.1$ of the rest mass of a spinning black hole can be extracted in this manner and can power relativistic jets.  Recent non-axisymmetric, relativistic MHD simulations have confirmed that this process can operate efficiently and an accreting, spinning black hole can be slowed down by electromagnetic torque and actually lose mass \cite{McKinney2012MNRAS.423.3083M}. (They also demonstrate that the jets that are formed are remarkably resilient and persistent.) A spinning black hole associated with a billion solar mass black hole and a quasar like 3C273 can produce an EMF of $\sim1$~ZV, sufficient for UHECR, and a power $\sim3\times10^{39}$~W.
\subsection{Diffusive Shock Acceleration}
\label{subsec:dsa}
A quite different approach due particle acceleration builds upon the original insight of Fermi\footnote{who used to vacation here in San Vito} \cite{Fermi1949PhysRev.75.1169} and others that posited that the energy gains are stochastic in relatively small steps over long intervals. Originally the particles were thought to gain energy by bouncing elastically of magnetized gas clouds. Nowadays, the clouds are replaced by hydromagnetic waves which can scatter resonantly when their wavelength matches the particle gyro frequency (see, e.g., \cite{kulsrud_plasma_2005}).  A variation on this idea is possible at a strong shock front where, in the simplest version, the wave speed relative to the fluid vanishes and it is the relative velocity between the scatterers on either side of the shock front that leads to efficient acceleration (\cite{Axford1977ICRC...11..132A, Krymskii1977DoSSR.234.1306K, Bell1978MNRAS.182..147B, Blandford1978ApJ...221L..29B}; or \cite{longair_high_2011} for a review). In the simplest version, low energy particles injected into the process are transmitted downstream with a power law in the momentum space distribution function with logarithmic slope of $-q=-3r/(r-1)$ where $r$ is the compression ratio. A value of $r\sim3.5$ appropriate for a fairly high Mach number and a monatomic gas, gives $q\sim4.2$ and a slope of the source spectrum $S(E)$ $\sim2.2$. This is just what is needed to account for the injection spectrum of Galactic supernova remnants.

The details, however, are both more complicated and more interesting (e.g. \cite{MalkovDrury2001RPPh...64..429M}).  As long as the cosmic rays are only a minor perturbation on the behavior of the background plasma, the transmitted spectrum will be proportional to the rate at which a few thermal particles are \emph{injected} into the acceleration process. However, there has to be some feedback  limiting this injection and the high efficiency suggests that the cosmic rays are part of the shock front. They have to be included in the Rankine-Hugoniot jump conditions among with the thermal plasma and, perhaps, the magnetic field. This has two consequences. The cosmic ray pressure decelerates the upstream flow and reduces $r$; their reduced pressure to energy density ratio increases $r$. Although the linear result is independent of the details of the diffusion coefficient, this is not true under actual nonlinear conditions. We know by comparing the observed thickness of shock fronts with Galactic cosmic ray propagation that the scattering frequency must be very much larger in the vicinity of the shock front than in the general interstellar medium and it is widely believed that the scattering waves are created by the cosmic rays themselves in their hurry to escape. The manner in which this feedback controls the injection process for protons, heavier nuclei and electrons\footnote{A very interesting possibility is that heavy nuclei are injected as charged grains which have large Larmor radii and that these grains are quickly broken down to their constituent atoms nuclei when they are accelerated to high speed.} is still poorly understood though impressive kinetic simulations are advancing our understanding of the problem and do demonstrate that the necessary suprathermal particles are generally present \cite{Caprioli2014ApJ...783...91C, Caprioli2014arXiv1401.7679C}.

Three types of linear instability leading to the growth of waves have been discussed (e.g., \cite{Schure2012SSRv..173..491S, Bykov2013SSRv..178..201B}). Firstly, the ``Bell'' mechanism relies upon the observation that the positively-charged cosmic ray protons are more plentiful  than the cosmic ray electrons and so as they stream through the background plasma they must be accompanied by a current of neutralizing electrons. This current can excite waves with wavelengths shorter than the Larmor radii $r_L$ of the proton cosmic rays. Secondly, a distribution of cosmic ray protons streaming faster, on  average, than the Alfv\'en speed, is unstable and will lead to the resonant growth of resonant Alfv\'en waves with wavelength similar to the proton Larmor radii. Thirdly, the cosmic ray pressure is likely to be anisotropic and when the component along the field exceeds that perpendicular to the field by more than the magnetic stress, hydromagnetic waves with energy larger than the cosmic ray Larmor radius can grow. This is known as the ``firehose'' instability. Other instabilities, notably the ``Weibel'' instability are thought to be relevant. The situation is, of course, even more complicated than this as the wave amplitudes are likely to be so large that wave-wave interactions are important and strong turbulence spectrum is established. The observation that cosmic rays also amplify magnetic field requires nonlinearity. A common assumption is that the effective diffusion coefficient for a cosmic ray of energy $E$ is as large as the ``Bohm'' value $\sim cr_L(E)\propto E$.

It is therefore quite likely that the highest energy cosmic rays stream furthest ahead of the shock front. Put another way, the first dynamically important warning that interstellar gas receives of an approaching strong shock front is a stream of escaping $\sim100$~TeV cosmic rays. Their distribution function is highly anisotropic and inverted in energy and, consequently unstable. The highest energy cosmic rays are those whose escape is impeded so that their mean escape speed at double the shock radius is roughly the shock speed. As their pressure will exceed the ambient interstellar pressure, they alone can create much of the magnetic field which eventually scatters lower energy cosmic rays closer to the shock front. This ``magnetic bootstrap'' mechanism \cite{BlandfordFunk2007AIPC..921...62B} ensures that many of the highest energy particles at all phases of the evolution of the supernova remnant will escape upstream and will avoid losing energy to expansion loss which is the fate of the lower energy cosmic rays transmitted downstream. The escaping particles therefore are disproportionately prominent in the total spectrum of cosmic rays accelerated over the complete lifetime of a remnant. Computing this spectrum, including all the many processes involved remains a challenge.

Relativistic shock fronts can also accelerate high energy particles though the physical processes involved are rather different from those present in diffusive shock acceleration. Numerical simulations are also advancing our understanding of the mechanisms involved and have great potential (e.g., \cite{Spitkovsky2008ApJ...682L...5S, Sironi2011ApJ...726...75S}).
\subsection{Reconnection}
\label{subsec:reconnection}
Magnetic field is known to undergo reconnection\footnote{``Reconnection'' is often a misnomer because the field lines that exchange partners had not previously met!} in laboratory and space plasma (e.g., \cite{priest_magnetic_2000}). The typical circumstance involves the juxtaposition of oppositely directed magnetic field in a highly conducting medium that create sufficiently strong gradients form at ``X'' points that allow the field lines to establish a quadrupolar ``flow'' through the medium near the X-point. The magnetic field is generally three dimensional and a component of field perpendicular to the flow can be convected  with it. Reconnection typically happens at current sheets which break up into parallel X-lines separated by O-like ``islands''. The detailed plasma physics of reconnection is quite complex involving strong localized currents that are subject to ``anomalous'' resistivity, Hall effects  and slow shock fronts. A reconnection site inevitably involves dissipation and it is a natural place for particle acceleration to take place and this is observed, for example in the Earth's magnetotail. However, the efficiency is generally quite low and the rate of dissipation of magnetic energy is slow.

New possibilities arise when the plasma is relativistic and these have been investigated in recent publications (e.g., \cite{Cerutti2014ApJ...782..104C, Sironi2014ApJ...783L..21S}). Hall effects are absent in a pair plasma. Electrodynamical integrations of the equations of motion in model reconnecting electromagnetic field demonstrate that test particles can be accelerated to impressively high energy especially along straight, non-radiating trajectories. Kinetic calculations show that the situation is more complex with islands forming which can scatter the particles and limit their energy gain. Relativistic reconnection remains an attractive but unproved possibility of acceleration of the highest energy particles though the efficiency with which it is likely to operate is unknown.

\subsection{Magnetoluminescence}
\label{subsec:maglum}
As we have described, observations of PWN, (notably the Crab Nebula), AGN, GRB and other high energy sources exhibit broad nonthermal continuum radiation emitted by freshly accelerated electrons and, probably, positrons. Sometimes the particle acceleration is extremely rapid and concentrated as demonstrated by rapid variability. Although the details remain highly controversial, it is generally acknowledged that the ``prime movers'' in these objects are spinning magnetized conductors and a large part of their power is carried off in the form of  a large scale electromagnetic wind comprising organized toroidal magnetic field accompanied by poloidal electric field. There will be high spatial frequency variation in this field, for example the stripes in the Crab wind and the consequences of instability. The whole flow will be permeated with electrical current. The environment is crucially different from that associated with stellar and disk coronae because the field is not anchored in a heavy conductor.

The power is presumably mostly electromagnetic close to the source. However, most of the emission comes from large distance and by this time the power will have been at least partly converted into relativistic plasma and ultimately into radiation. The problem is to understand just how this happens. Several schemes have been well-explored. Relativistic shocks are commonly invoked and are surely present. However, when the magnetic field is dominant, the shocks are weak and relatively ineffectual accelerators. Another option is that a ``top-down'' inertial range turbulence develops with the energy flow feeding the particle acceleration on short scales. The hydromagnetic waves may be damped by gyroresonant absorption when the absorption rate is faster than the eddy turnover timescale. As with fluid turbulence there is likely to be some intermittency to account for the rapid variability. However, the challenge to turbulence models is to accelerate to a broad range of particle energies efficiently. A third possibility is relativistic reconnection which we have just introduced. Here, the reconfiguration of the magnetic field leads to a loss of magnetic energy increasing the particle energy. In relativistic reconnection the relevant scaling speed is the speed of light as opposed to the Alfv\'en speed. However the reconnection takes place in a series of line currents and so there is a bottleneck present which will lower the rate at which there can be a large, volumetric changes, typically by a logarithm $\sim10$. It has not been demonstrated that it can process large volumes of magnetic energy at relativistic speed.

These considerations motivate a different approach. We propose that the large loops of toroidal field undergo a relatively slow reconnection to create shorter flux tubes that contract under magnetic tension limited by the inevitable tangling of the field lines. The resultant magnetic ``cells'' then evolve initially slowly failing catastrophically at the end producing large, inductive electric field over an extended volume and accelerating high energy electrons. If the magnetic energy is radiated efficiently, the pressure support will be lost and the external pressure will cause an implosion of the emitting volume and perhaps further emission.  We call the general process ``magnetoluminescence''\footnote{By analogy with ``sonoluminescence'' (e.g. \cite{Putterman10.1146/annurev.fluid.32.1.445}) where fluid irradiated with ultrasound can cavitate and the resulting bubbles implode, producing ultraviolet and X-ray radiation.}.

The behavior of magnetic field under natural and laboratory conditions is a constant source of surprise. Locally, we can decompose the current density associated with a specific magnetic field into a parallel and a perpendicular component.
The former exerts no stress and if that is all that is present, then the field is said to be ``force-free''. The parallel current causes the magnetic field lines to develop a twist.  When there is a perpendicular current and field varies slowly on the scale of the particle Larmor radii, so that the pressure tensor just contains a component  $P_{||}$ along the field and a component $P_\perp$ perpendicular to the field, then the equation of motion can be used to show that
\begin{equation*}
\begin{split}
\vec j_\perp=\frac{\vec B\times(\vec B\cdot\nabla)\vec B}{B^4}P_{||}-\left[\nabla\times\left(\frac{P_\perp\vec B}{B^2}\right)\right]_\perp+\frac{\vec B\times\nabla B}{B^3}P_\perp\\
+\frac{\vec E\times\vec B}{B^2}\nabla\cdot E+\frac{(\vec g-\vec a)\times\vec B}{B^2}\rho.
\end{split}
\end{equation*}
The terms on the right hand side are associated with the curvature, perpendicular magnetization, gradient, $\vec E\times\vec B$, gravitational and acceleration $\vec a$ drifts respectively and these occur in the absence of collisions (discussions on particle motions in electromagnetic field can be found in, e.g., \cite{RossiOlbert1970ips..book.....R}). It is the resistance associated with $\vec j_{||}$ that seems to be most important for converting magnetic energy into particle energy. This description breaks down if and when the particle gyro radii approach the magnetic scale length and there could be a sort of phase transition to chaotic trajectories.

A second characterization of magnetic field is a little harder to express. It is best illustrated by considering a model force-free magnetic structure called a ``spheromak'' (e.g., \cite{rosenbluth_mhd_1979, bellan_spheromaks_2000}). If, for simplicity, we suppose the magnetic cell is bounded by a rigid spherical conductor we can find eigenfunction solutions of the relevant (Helmholtz) equation inside the conductor, characterized by ``quantum'' numbers $n,l,m,s$ that are similar to their counterparts in the solution to the hydrogen atom\footnote{Considerations analogous to these featured in nineteenth century attempts to describe atoms but today are most useful in highlighting the essential differences between classical and quantum mechanics.}. When we analyze the magnetic field lines in these solutions, we find that some of them are confined to two dimensional surfaces while others appear to wander ergodically over three dimensional space. The two dimensional field lines comprise a ``rope''. Ropes are twisted perhaps hierarchically, like real ropes because they carry field-aligned current. This gives them integrity and allows them to be distinguished inside a larger volume of wandering field that carries less current.  In reality, as opposed to in applied mathematical solutions, magnetic ropes are also likely to be ``hairy'' with minor, dendritic field lines leaving and joining the rope surface. Magnetic ``ropes'' in spheromak solutions can be linked and knotted and it is possible to imagine transitions between steady state solutions that preserve the major features of the flux linkage and topology, while changing the energy. The linkage is often measured by the magnetic helicity $\int dV\vec A\cdot\vec B$ which may be preserved in ``permitted'' transitions. Changes like this seem to persist when the boundary conditions are changed, specifically if they are isobaric instead of isochoric and when plasma pressure is included. (Some related discussions on magnetic field topology can be found in e.g., \cite{Moffatt1978mfge.book.....M, Parker1979cmft.book.....P}.)
\begin{figure}[htb]
  \centering
  \includegraphics[width=0.3\textwidth]{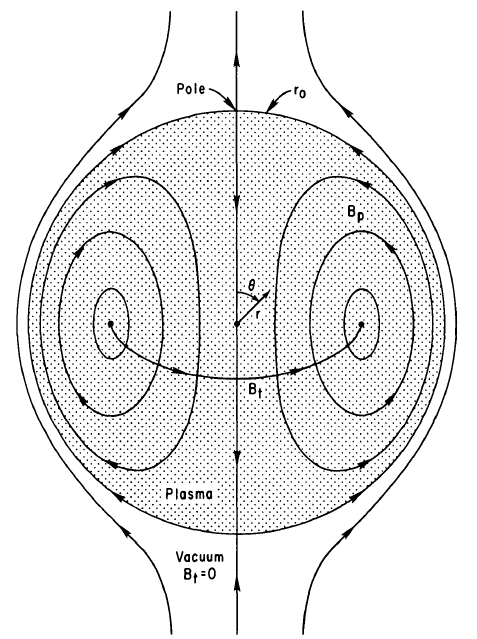}\\
  \caption{Classical spheromak, namely the n=1, l=1, m=0 solution (from \cite{rosenbluth_mhd_1979}).}\label{fig:spheromak}
\end{figure}

We should distinguish two important cases from a topological perspective. The active cell may contain one or more ``open'' magnetic rope segments which we call a braid introducing the name ``unbraid'' for the case when the segments are essentially parallel lines. Alternatively, the ropes may be closed into one or more ``knots'' including the ``unknot'' which is topologically equivalent to a circle. Braids and knots may be linked in a single cell. A specific topological structure may be created in a more complex configuration containing metastable, tangled states that can suddenly disentangle. For example a single magnetic rope can form a slip knot, a truckers' hitch or a sheepshank. Not without ambiguity, we call this a ``hitch''. Our basic contention is that topology is changed slowly through reconnection in response to slow large scale motions and suddenly through unravelling hitches. While the former can lead to steady particle acceleration and continuum emission, it is the sudden disentangling of hitches that leads to higher energies and rapid variability.

\begin{figure}[htb]
  \includegraphics[width=0.45\textwidth]{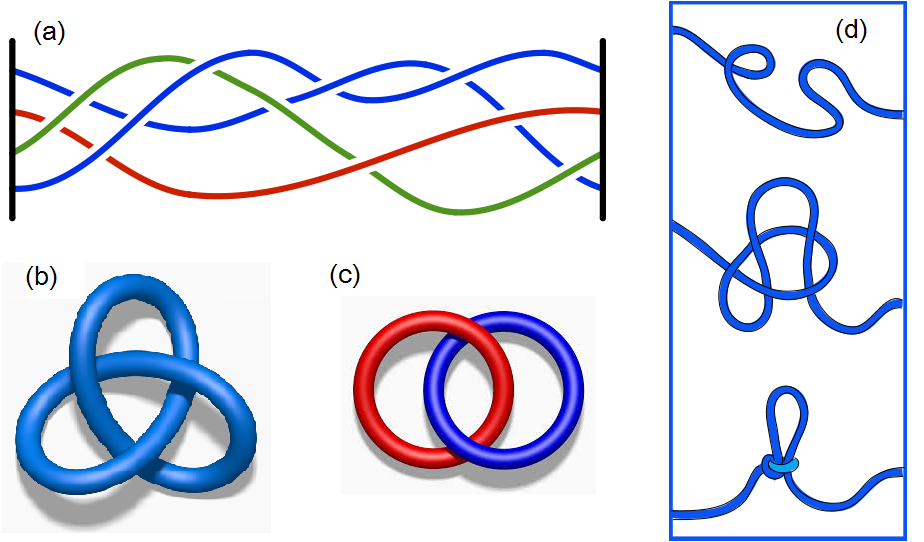}\\
  \caption{An illustration of topologies the magnetic ``ropes'' may take. (a) Braid. (b) Knot. (c) Link. (d) Hitch, which is topologically equivalent to a straight line.}\label{fig:knots}
\end{figure}

The actual mechanism of particle acceleration introduces several fascinating possibilities. If there are electrons and positrons counter-streaming along the field then they can excite electrostatic and electromagnetic waves which may scatter the particles by changing their pitch angles and be responsible for the resistance. There will be a non-zero $\vec E\cdot\vec B$, which can accelerate the particles at the expense of the magnetic energy. How the particle momentum space distribution function  evolves is very difficult to guess without detailed kinetic calculations, but there is at least the possibility of a ``runaway'' developing with most of the energy going into the fastest low pitch angle particles. When the magnetic ropes are closed it is also possible that the particles perform many circuits slowly gaining energy as in a storage ring. Alternatively, when the hitches disentangle, there should be large, inductive electric field which can lead to a rapidly changing electromagnetic geometry with chaotic particle motion. In the limit that $E$ approaches $cB$, the acceleration can become radiation-reaction limited and the effective resistivity can be due to the virtual photons associated with the curving magnetic field. In this limit the synchrotron photons have energies $\sim\alpha^{-1}m_ec^2$ close to what is observed in the most dramatic Crab Nebula flares.

Clearly, kinetic simulations will be necessary to understand these various possibilities. Further discussion will be presented elsewhere.

\section{Future Possibilities}
\label{sec:future}
The future prospects in the study of cosmic rays and their origin are bright. First and foremost the precision with which the various spectra are measured will continue to improve even using existing facilities such as AMS. Features in these spectra at a specific rigidity (momentum per unit charge) could be found in different species and might provide strong clues concerning the propagation. UHECR will continue to be detected by Auger(s) and the Telescope Array. Their high energy spectra will be better measured and their composition determined. Further Voyager measurements may help us define the interstellar cosmic ray spectrum much better. When we consider the more general problem of high energy astrophysical particle acceleration, which is at least mechanistically relevant to cosmic ray origin, then several new capabilities are anticipated. VHE neutrino astronomy is about to be born presuming that we will eventually be able to find the sources of the cosmic neutrinos that are being detected. The reality of FRBs should be settled and if they are cosmic identifications should come soon.

Advanced LIGO followed by other gravitational radiation interferometers should be on the air in a couple of years and are expected to starting seeing merging neutron stars and perhaps other sources soon \cite{LIGO2013arXiv1304.0670L}. It is hard to imagine that any source of detectable gravitational radiation (even merging black holes) avoids particle acceleration. Another approach to radiation detection that might even beat LIGO is using an array of roughly a hundred millisecond pulsars --- the International Pulsar Timing Array --- to search for a background of low frequency waves from merging black holes in AGN and other more speculative sources \cite{Hobbs2010CQGra..27h4013H}. The ALMA mm/submm telescope is becoming fully operational and can be phased up to operate with other mm telescopes to form a powerful VLBI array --- the Event Horizon Telescope --- that is just capable of resolving the central black holes at the Galactic Center and in M87 \cite{Doeleman2010evn..confE..53D, Lu2014ApJ...788..120L}. TeV gamma ray astronomy will also develop with the construction of the Cerenkov Telescope Array intended to cover the whole sky from the north and the south \cite{CTA2014arXiv1405.5696V}. Note that most of these sources are transients so that  identification and enlightenment comes from ``multi-messenger'' investigations.

Fluid and kinetic simulations of the many astrophysical sites introduced in this review already have the capability, fidelity and dynamic range to perform experimental investigation and answer questions that are unreachable by analysis. They are no less important to the rapidly developing field of High Energy Density physics where powerful lasers and giant capacitor banks are deployed to create relativistic plasma by compressing matter under extreme conditions. Recent experiments have been marked by greatly improved diagnostics  which are enabling a much better understanding of how plasmas behave in practice.

It is hoped that this sketchy overview of astrophysical particle acceleration, especially as applied to the hundred year old mystery of cosmic ray origin, will introduce the more detailed examination of these topics in the following talks.
\section*{Acknowledgements}
RB thanks the organizers for the opportunity to attend this meeting and many attendees for helpful discussions. Tom Abel, Jon Arons, Rolf Buehler, Stefan Funk, Hui Li, Mikhail Malkov, Igor Moskalenko, Angela Olinto, Ellen Zweibel and the Fermi team are also thanked for collaboration and discussion of many of the topics discussed above. Support by the Miller and Simons Foundations (to RDB) and under NSF grant AST1212195 and DOE contract DE-AC02-76SF00515 as well as the KIPAC Fellowship made possible by the Kavli Foundation are gratefully acknowledged.




\nocite{*}
\bibliographystyle{elsarticle-num}
\bibliography{sanvito}








\end{document}